\documentclass[a4paper, preprint,12pt,sort&compress]{elsarticle}

\usepackage{epsfig}

\usepackage[dvipdfmx]{color}

\usepackage{times} 

\usepackage{amssymb}
\usepackage{lineno}

\usepackage[utf8]{inputenc} 
\usepackage[T1]{fontenc} 
\usepackage{mathptmx} 

\journal{Journal of Magnetism and Magnetic Materials}

\begin{document}
\begin{frontmatter}

\title{Magnetocaloric particles of the Laves phase compound HoAl$_{2}$
prepared by electrode induction melting gas atomization}

\author[1]{Takafumi D. Yamamoto\corref{cor1}}
\ead{YAMAMOTO.Takafumi@nims.go.jp}
\author[1]{Hiroyuki Takeya}
\author[1]{Akiko T. Saito}
\author[1]{Kensei Terashima}
\author[1]{Takenori Numazawa}
\author[1,2]{Yoshihiko Takano}

\cortext[cor1]{Correpsonding author}

\address[1]{National Institute for Materials Science, Tsukuba, Ibaraki 305-0047, Japan}
\address[2]{University of Tsukuba, Tsukuba, Ibaraki 305-8577, Japan}

\begin{abstract}
Processing magnetocaloric materials into magnetic refrigerant particles is
an essential issue in developing high-performance magnetic refrigerators.
Here, we succeed in stably producing magnetocaloric particles of the promising material HoAl$_{2}$
by a newly devised method based on electrode induction melting gas atomization process.
The particle size range is on the order of submillimeters,
which is suitable for practical refrigeration systems.
The resulting particles with less contamination have
good morphological, magnetic, and magnetocaloric properties:
(i) almost spherical shapes with few internal pores,
(ii) a sharp ferromagnetic transition around 30 K,
and (iii) a large magnetocaloric effect comparable to the bulk counterpart.
These features suggest the HoAl$_{2}$ gas-atomized particles have
the potential of use as a magnetic refrigerant.
The presented method can be applied not only to HoAl$_{2}$
but also to other brittle magnetocaloric materials with high melting points,
facilitating the production of various magnetic refrigerants
needed to develop magnetic refrigerators for hydrogen liquefaction.
\end{abstract}

\begin{keyword}
Magnetocaloric effect,
Intermetallic compounds,
Gas-atomization,
Thermodynamic properties
\end{keyword}

\end{frontmatter}
\newpage
\section{Introduction}
Magnetocaloric effect (MCE) is a magneto-thermodynamic phenomenon
in which a magnetic material absorbs/generates heat
when exposed to a varying magnetic field.
The magnetic refrigeration based on the phenomenon has attracted attention
as a promising alternative to conventional gas compression/expansion refrigerations,
owing to the advantages of being highly efficient, environmentally friendly,
and feasible over a wide temperature range from extremely low temperatures to room temperature
\cite{Gschneidner-RPP-2005,Kitanovski-Text-2015,Lyubina-JPD-2017,Franco-PMS-2018}.
Recently, the need for lowering any costs related with liquid hydrogen,
a potential medium for economical transportation and storage
\cite{Sherif-IJHE-1997, Wijayanta-IJHE-2019},
has increased toward realizing a future society using hydrogen energy.
In this context, great interest has been focused on
the development of high-performance magnetic refrigeration systems for hydrogen liquefaction
\cite{Numazawa-Cryo-2014,Zhang-PhysicaB-2019},
which has driven extensive research on
exploring for new magnetocaloric materials working at cryogenic temperatures
\cite{Zhang-PhysicaB-2019,Li-SSC-2014,Matsumoto-JMMM-2017,Omote-Cryogenics-2019,Pedro-NPG-2020}
and developing highly efficient refrigerators based on active magnetic regenerator (AMR) cycles
\cite{Utaki-Cryoc-2007,Matsumoto-JPCS-2009,Kim-Cryogenics-2013}.

The Laves phase compounds RT$_2$ (R = Gd-Yb, T = Al, Ni, Co) exhibit
large MCEs associated with a ferromagnetic transition
occurring in the temperature range between
the two liquefaction temperatures of hydrogen (20 K) and nitrogen (77 K)
\cite{Hashimoto-ACEM-1986,Tomokiyo-ACE-1986,Zhu-Cryogenics-2011}.
Moreover, their Curie temperature ($T_{\rm C}$) is easy to tune by element substitutions.
Such features make these compounds to be one of the candidate materials
for magnetic refrigerants used in magnetic liquefiers for hydrogen.
On the other hand, practical use of promising magnetocaloric materials in AMR systems
requires them to be processed into spherical particles with a size of 200-400 $\mu$m
for optimal system performance
\cite{Barclay-CPE-1984,Yu-IJR-2010}.

The material properties of the RT$_{2}$ compounds may hinder producing spheres
by some conventional spheroidization processes.
For example, they have high melting points above 1000 $^{\circ}$C and high reactivity.
In the processes that melt and/or superheat a target material in a crucible,
such as the centrifugal atomization method
\cite{Osborne-ACE-1994,Matsumoto-JPconf-2012,Wolf-PowderTech-2020}
and the gas atomization process
\cite{Anderson-Proceed-1992,Aprigliano-ACE-1992},
these properties may cause the reaction of the target material and the crucible,
preventing the mass production of high-quality particles.
The plasma rotating electrode process
\cite{Miller-Cryoc-2001,Fujita-JJAP-2007}
can produce high-purity spherical particles
by non-contact melting of a material's electrode rod that rotates at high speed.
Nevertheless, the brittle nature of RT$_{2}$ compounds can make it challenging
to fabricate electrode rods that can withstand for high-speed rotations.

As a viable route for producing magnetic refrigerant particles,
we have focused on the electrode induction melting gas atomization (EIGA) technique
\cite{Antony-JOM-2003},
in which an electrode rod of a target material is inductively melted in a non-contact manner.
Very recently, we actually succeeded in fabricating spherical particles of
a giant magnetocaloric material, HoB$_{2}$, with a melting point of 2350 $^{\circ}$C
\cite{TDY-APA-2021}.
In the EIGA process, ingots are typically used as electrode rods.
However, since the RT$_{2}$ compounds are pretty sensitive to thermal shock,
the breakdown of the ingot can occur during induction heating.
In this study, we devise a new method to use a sintered body
as the electrode rod in the EIGA process
and realize stable production of spherical particles of HoAl$_{2}$,
one of the RT$_{2}$ compounds with the $T_{\rm C}$ of about 30 K
\cite{Hashimoto-ACEM-1986}
and a melting point of 1530 $^{\circ}$C.
The resulting particles are evaluated in terms of
morphological, crystallographic, and physical properties.
%
%
\section{Material and methods}
\begin{figure}[t]
\centering
\includegraphics[width=130mm]{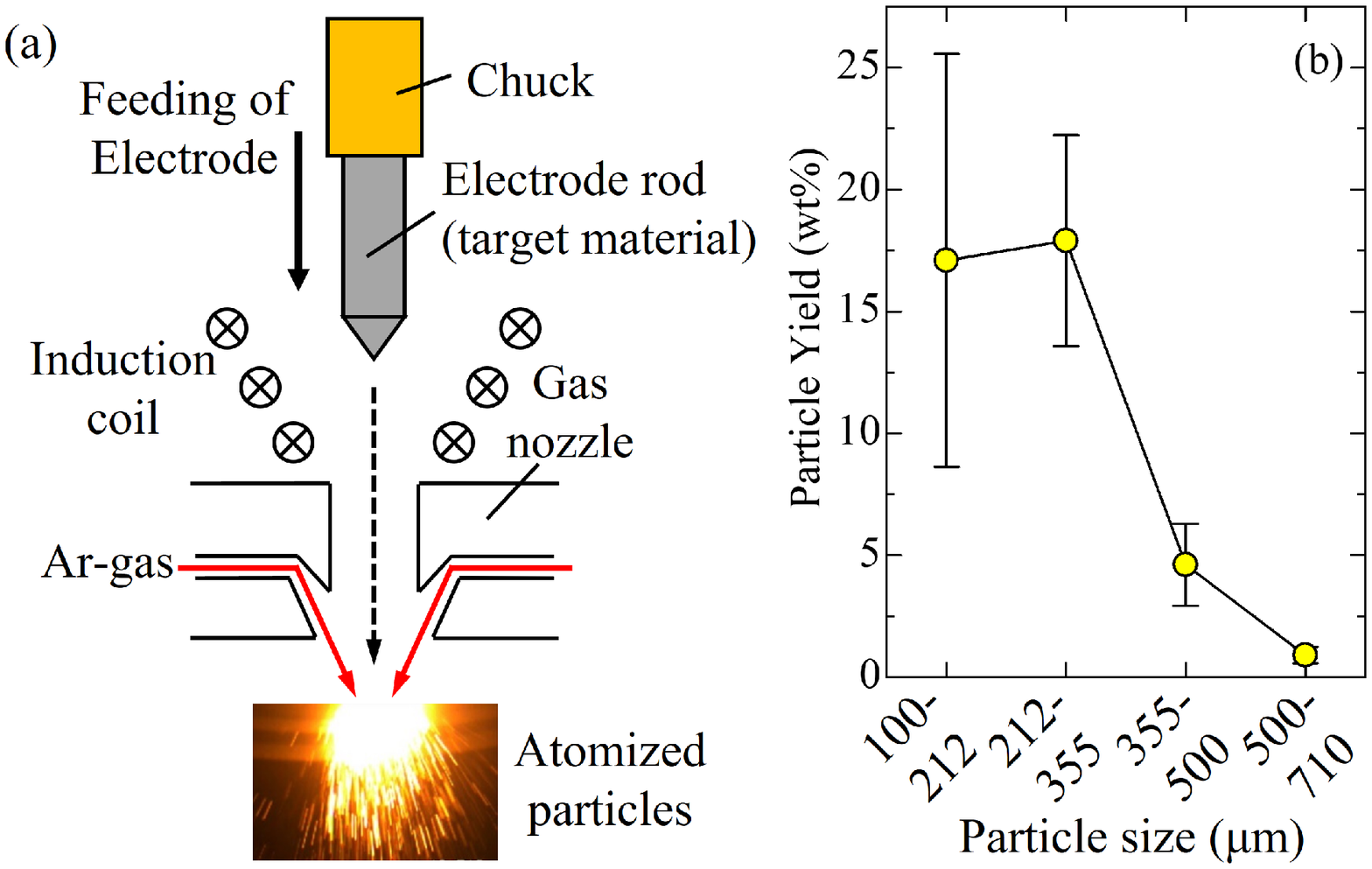}
\caption{(Color Online) (a) Schematic picture of the EIGA process.
(b) Particle size distribution of HoAl$_{2}$ gas-atomized particles.}
\label{fig:EIGA}
\end{figure}

Master alloy of HoAl$_{2}$ was prepared by high-frequency vacuum melting
from a stoichiometric mixture of Ho (3N) and Al (4N) elements.
The alloy was crushed into powder, filled into a quartz tube,
and sintered in a vacuum at 1000-1200 $^{\circ}$C for 20-24 h.
The resulting sintered rods were 400-500 g in weight for each.
Figure \ref{fig:EIGA}(a) depicts a schematic image of the present EIGA process.
In this process, one end of the sintered rod was fixed with a chuck and hung,
and the other end was inductively melted by an induction coil in an Ar-atmosphere.
The molten metal freely fell into the lower sample chamber,
atomized into droplets by jetting Ar-gas,
and then solidified into particles while flying in the chamber.

After the EIGA process, the collected atomized powder was sieved into six fractions,
$<$100 $\mu$m, 100-212 $\mu$m, 212-355 $\mu$m, 355-500 $\mu$m, 500-710 $\mu$m, and $>$710 $\mu$m,
through a series of JIS Z 8801 standard sieves with a FRITSCH ANALYSETTE 3 vibratory sieve shaker.
Since the sieved powder was composed of round particles and irregular ones,
the former were manually separated from the latter by rolling them on a sloping board,
though the particles less than 100 $\mu$m in diameter could not be separated well.
No heat treatment was not applied to both the master alloy and the atomized particles.

The particle morphology was observed using
a Hitachi SU-70 scanning electron microscope (SEM) operated at 20 kV.
The energy-dispersive X-ray spectroscopy (EDS) analysis was performed
to observe element distribution of the particles using a Hitachi TM4000 operated at 10 kV.
The particles were embedded in bakelite resin with carbon filler
for polishing to a flat surface prior to cross-section observation.
Powder X-ray diffraction (XRD) measurements were carried out at room temperature
by a Rigaku MiniFlex600 diffractometer (Cu K$\alpha$ radiation).

The two magnetocaloric properties, namely,
the magnetic entropy change $\Delta S_{\rm M}$
and the adiabatic temperature change$\Delta T_{\rm ad}$
were evaluated from the magnetization and specific heat data
for the round particles and the master alloy.
The details of our evaluation method for $\Delta S_{\rm M}$ and $\Delta T_{\rm ad}$ were
described elsewhere \cite{TDY-JPD-2019,TDY-JMMM-2020}.
Temperature dependence of magnetization ($M$-$T$ curve) was measured
by a Quantum Design SQUID magnetometer
from 60 to 2 K under various magnetic fields ($\mu_{0} H$) ranging from 0.01 to 5 T.
Field dependence of magnetization ($M$-$H$ curve) was taken at 2 K between 0 T and 5 T.
Specific heat ($C$) was measured at 0 T between 2 K and 65 K
by a thermal relaxation method using a Quantum Design PPMS.
In the magnetization measurements,
the master alloy had a rectangular shape with dimensions of 1.5 $\times$ 0.3 $\times$ 0.3 mm$^{3}$,
and dozens of the round particles were arranged vertically long
in contact with each other with an aspect ratio of 2-4.
The magnetic field was applied along the longitudinal direction of
both the shaped master alloy and the arranged particles.
The magnetization data in this study were not corrected for the demagnetization effect
because it is difficult to determine the exact demagnetization factor for the arranged particles.
%
\section{Results and discussion}
\subsection{Particle size distribution}
Figure \ref{fig:EIGA}(b) shows the yield of HoAl$_{2}$ gas-atomized particles
as a function of particle size.
Here the yield was calculated as the ratio of
the weight of the round particles for each diameter
to the total weight of the collected atomized powder.
The present EIGA process was able to run many times,
so the average values and the corresponding standard deviations are presented together.
The particle size is distributed in the range of 100-710 $\mu$m,
roughly centered around 212-355 $\mu$m,
for which the yield of 14-22 wt\% is obtained.
This result contrasts with that the particle size is typically 100 $\mu$m or less
in conventional EIGA experiments
\cite{Franz-Titanium-2008,M_Wei-Vaccum-2017,Xie-IOPconf-2019,Guo-ActaMetall-2017,Sun-IOPconf-2019}.
The atomizing gas pressure values in this study, 1.5-3.5 MPa, are lower than
those used in these previous studies.
The low gas pressure means that the kinetic energy of the atomizing gas is low,
which gives the molten metal a moderate breakup effect
\cite{Xie-IOPconf-2019}.
This should have provided the particle size range on the order of submillimeters
that meets the requirements of AMR systems.
\subsection{Morphological and microstructural properties}
\begin{figure}[t]
\centering
\includegraphics[width=120mm]{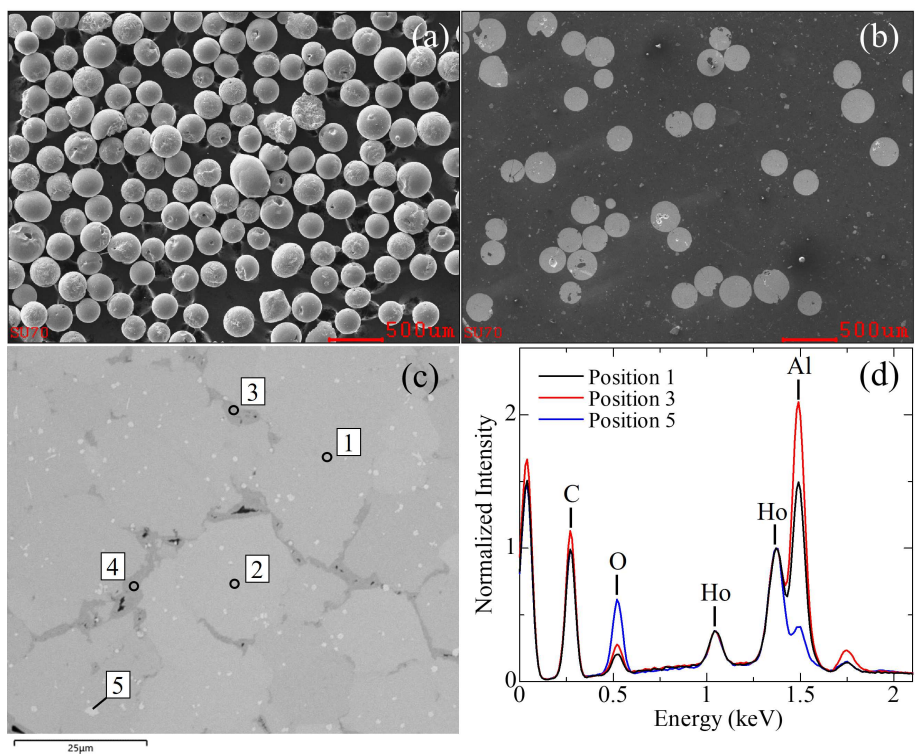}
\caption{(Color Online) SEM images of (a) surface and (b) cross section
for HoAl$_{2}$ particles with 212-355 $\mu$m diameter.
(c) A magnified cross-sectional backscattering electron image of the particle.
(d) EDS spectrum at representative potisions in (c).}
\label{fig:SEM}
\end{figure}

Figure \ref{fig:SEM}(a) shows an SEM image of
the surface for HoAl$_{2}$ particles with 212-355 $\mu$m diameter.
The majority of the particles are almost spherical,
but some are fractured or have small hollows on the surface.
The latter is probably due to particle/particle and particle/chamber collisions.
It is worth noting here that the particles have few internal pores,
as seen from the cross-sectional image in Fig. \ref{fig:SEM}(b).
The internal pore is a well-known significant issue
that causes degradation of gas-atomized particles
and results from the atomizing gas trapped inside the droplet.
These pores are more often found in particles larger than 100 $\mu$m
\cite{Guo-ActaMetall-2017},
whereas this is not the case in the HoAl$_{2}$ particles.
According to the consideration based on fluid mechanics
\cite{Sun-IOPconf-2019,Ternovoi-PowMetall-1985},
the higher the atomizing gas pressure, the higher the probability of gas-trapping.
Therefore, the low atomizing gas pressure is also likely to
help prevent the formation of internal pores.

Figures \ref{fig:SEM}(c) and \ref{fig:SEM}(d) show
a magnified cross-sectional backscattering electron image of the particle
and EDS spectrum taken at several positions.
One can find three characteristic areas:
the major light gray area (such as position 1 and 2),
the dark gray area (such as position 3 and 4),
and the white spot (such as position 5).
Compared to the light gray area,
the dark gray area has a stronger relative peak intensity of Al to Ho in the EDS spectrum,
implying that this area corresponds to HoAl$_{3}$.
In the white spot area, the intensity of Al is weak and that of O is strong,
indicating the presence of Ho$_{2}$O$_{3}$.
Note that the microcracks observed in the dark gray areas may have occurred
when the particles were polished.
\subsection{Crystallographic properties}
\begin{figure}[t]
\centering
\includegraphics[width=120mm]{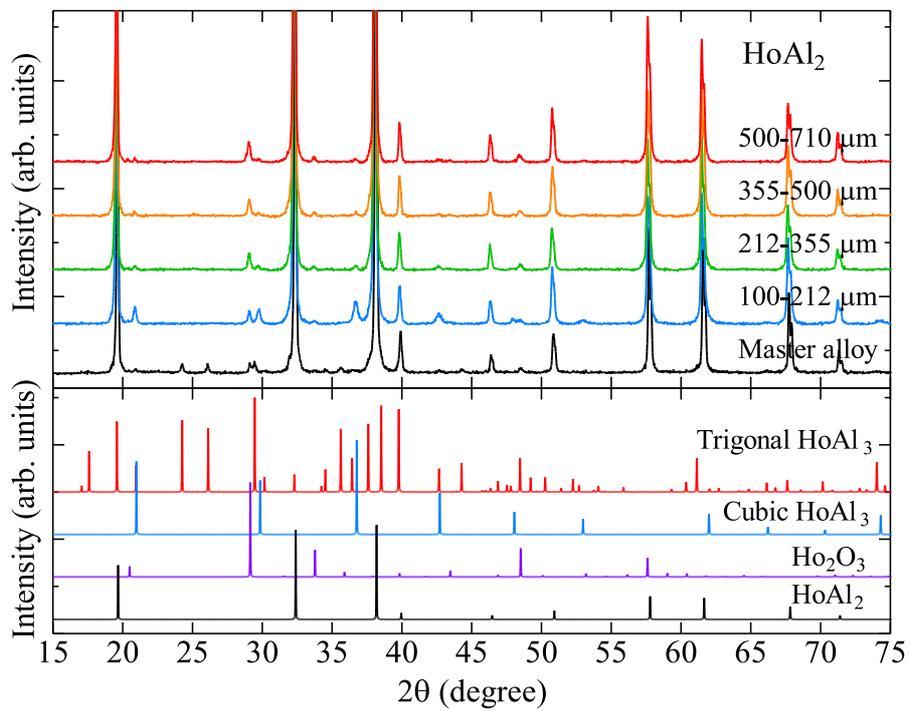}
\caption{(Color Online)
Upper panel: Powder XRD patterns of the master alloy of HoAl$_{2}$
and HoAl$_{2}$ particles with various diameters between 100 $\mu$m and 710 $\mu$m.
Lower panel: The calculated patterns of HoAl$_{2}$ \cite{Grossinger-JMMM-1976},
Ho$_{2}$O$_{3}$ \cite{Maslen-ActaCry-1996},
trigonal HoAl$_{3}$ \cite{Havinga-JLCM-1975},
and cubic HoAl$_{3}$ \cite{Cannon-JLCM-1975}
obtained with RIETAN-FP \cite{Izumi-SSP-2007}.}
\label{fig:XRD}
\end{figure}

The upper pannel of Fig. \ref{fig:XRD} shows
XRD patterns of HoAl$_{2}$ spherical particles
with various diameters between 100 $\mu$m and 710 $\mu$m.
The data for the master alloy is shown at the bottom for comparison.
The calculated patterns of some materials including HoAl$_{2}$
are shown in the lower panel.
The primary phase is indexed as HoAl$_{2}$ \cite{Grossinger-JMMM-1976}
in both the master alloy and the atomized particles.
Moreover, there are no differences in position and sharpness of diffraction peaks,
suggesting that the atomization process does not affect
the crystallographic properties of HoAl$_{2}$.
The intensity of impurity peaks does not increase much
before and after the atomization process,
which implies less contamination during a series of processes.
However, the type of impurities is different between the two samples;
in addition to Ho$_{2}$O$_{3}$ \cite{Maslen-ActaCry-1996}, which is common to both,
the master alloy has trigonal HoAl$_{3}$ \cite{Havinga-JLCM-1975},
and the atomized particles has cubic HoAl$_{3}$ \cite{Cannon-JLCM-1975}.

The intensity of Ho$_{2}$O$_{3}$ peaks tends to increase with increasing particle size.
In general, oxidization is facilitated at high temperatures.
Given that the larger particles need longer cooling times,
this result implies that the gas-atomized particles take up oxygen during solidification.
Note that the HoAl$_{2}$ particles do not oxidize under ambient conditions.
It is interesting that HoAl$_{3}$ changes its crystal structure through the atomization process.
Cannon and Hall \cite{Cannon-JLCM-1975} have reported that such transformation occurs
at high temperatures ($\sim$ 1000 $^{\circ}$C) and high pressures (6.4 GPa),
but this is far from the present experimental conditions.
One possibility is that the rapid solidification of the atomized particles causes the transformation.
Indeed, the cubic HoAl$_{3}$ peaks are prominent in smaller particles,
in which the solidification time may be shorter.
The experiments are underway to verify the quenching effect on trigonal HoAl$_{3}$.
\subsection{Effects of atomization process on the magnetism}
\begin{figure}[t]
\centering
\includegraphics[width=120mm]{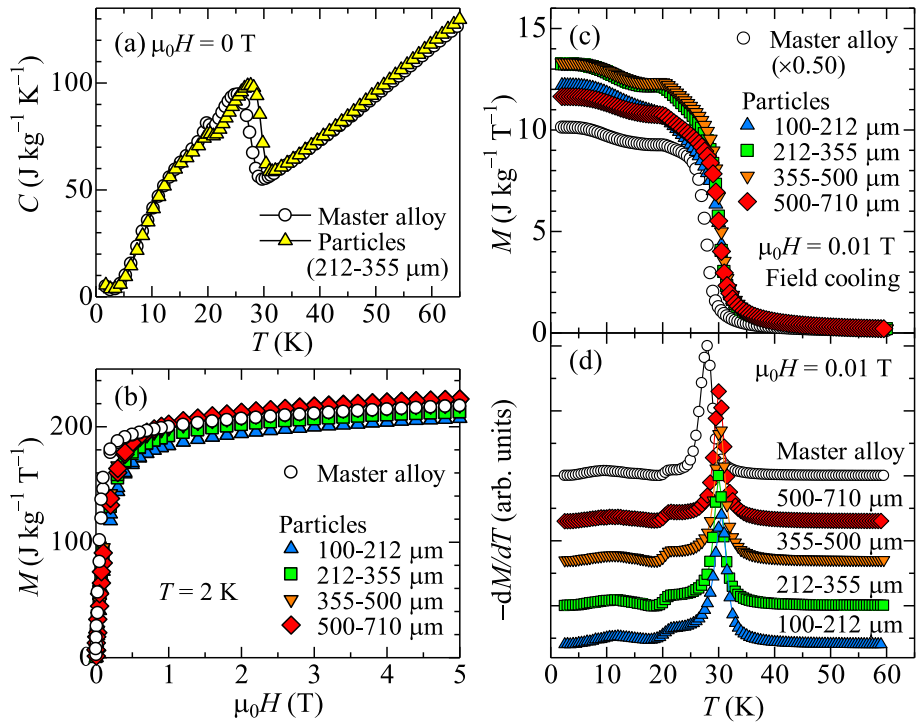}
\caption{(Color Online) (a) Temperature dependence of specific heat at zero magnetic field
in the master alloy and the HoAl$_{2}$ particles with 212-355 $\mu$m diameter.
(b) Magnetic field dependence of magnetization at 2 K
for the master alloy and the particles with various diameters.
(c) and (d) Magnetization and the temperature derivative of magnetization d$M$/d$T$
as a function of temperature at 0.01 T in field cooling processes for all the samples.}
\label{fig:Properties}
\end{figure}

Figure \ref{fig:Properties}(a) shows the temperature dependence of specific heat at 0 T
in HoAl$_{2}$ particles with 212-355 $\mu$m diameter and the master alloy.
Both samples exhibit a specific heat peak around 30 K,
ascribed to the paramagnetic-ferromagnetic transition in HoAl$_{2}$.
The peak in the particles is as clear as that in the master alloy,
suggesting that the ferromagnetic transition in the former is sharp as well as the latter.
In addition, as shown in Figs. \ref{fig:Properties}(b) and \ref{fig:Properties}(c),
the $M$-$H$ curves at 2 K and the $M$-$T$ curves under 0.01 T are
similar for both the atomized particles and the mater alloy.
These results suggest that the HoAl$_{2}$ gas-atomized particles have
good sample quality comparable to the bulk counterpart.
Note that the difference in magnitude of $M$ at low magnetic fields is
due to that in demagnetization factor between the master alloy and the particles.

The specific heat peak is shifted to higher temperatures in the atomized particles,
which is indicative of the increase of $T_{\rm C}$.
Here we define $T_{\rm C}$ as a peak temperature of
the temperature derivative of magnetization (d$M$/d$T$) at 0.01 T.
As can be seen from d$M$/d$T$ in Fig. \ref{fig:Properties}(c),
the $T_{\rm C}$ is evaluated to be 28 K for the master alloy
and 30 K for all the particles with various diameters.
Considering that the X-ray diffraction peaks of HoAl$_{2}$ does not change
before and after the atomization
and the $T_{\rm C}$ is independent of particle diameter, or equivalently, solidification time,
the structural properties are unlikely to be responsible for the $T_{\rm C}$ shift.
Rather, the difference in $T_{\rm C}$ between the atomized particles and the master alloy seems to be
attributed to whether the crystal structure of HoAl$_{3}$ is trigonal or cubic.
Trigonal HoAl$_{3}$ has been reported as an antiferromagnet
with a N$\acute{\rm e}$el temperature of 9 K \cite{Buschow-HFM-1980},
while the magnetism of cubic HoAl$_{3}$ has not been clarified.
The difference in their magnetic properties might affect
the physical properties of HoAl$_{2}$,
but the further investigations are needed.

One can find another anomaly around 20 K in both $C$ and $M$.
This is ascribed to a spin-reorientation transition
\cite{Hill-JSSC-1973,Barbara-PhysLett-1975},
in which the easy axis of magnetization changes
from <100> direction above 20 K to <110> direction below 20 K.
The anomaly looks similar in all the samples,
implying that this transition is largely unaffected by the atomization process.
This result seems to be consistent with the absence of changes
in the crystallographic properties of HoA$_{2}$ by the atomization process,
as strong magneto-elastic coupling has been observed
associated with the spin-reorientation transition
\cite{Ibarra-JPC-1988}.
\subsection{Magnetocaloric properties}
\begin{figure}[t]
\centering
\includegraphics[width=90mm]{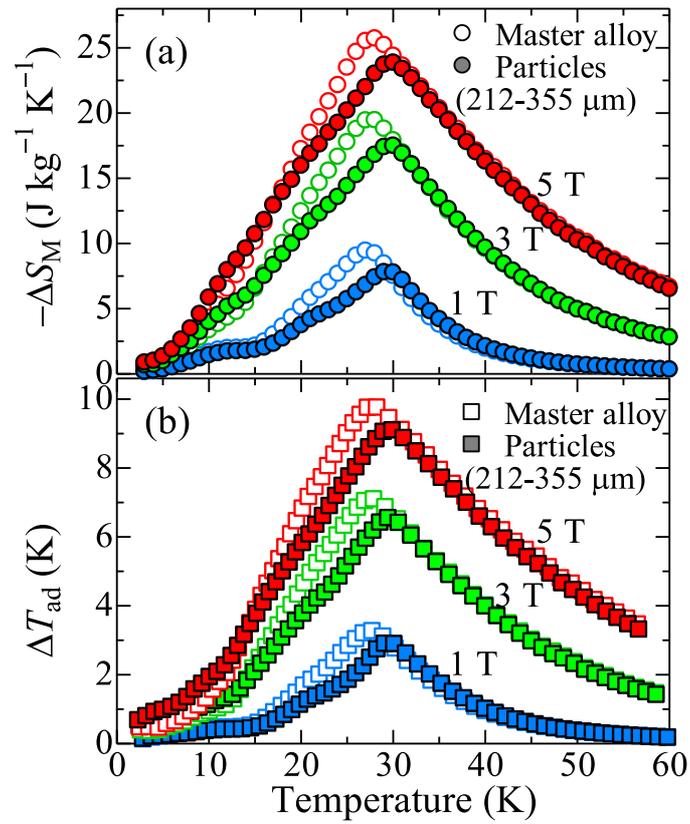}
\caption{(Color Online) (a) Magnetic entropy change and (b) adiabatic temperature change
for $\mu_{0} \Delta H$ = 1, 3, and 5 T
in the master alloy and HoAl$_{2}$ particles with 212-355 $\mu$m diameter.}
\label{fig:MCE}
\end{figure}

Here we will evaluate the magnetocaloric properties
for HoAl$_{2}$ spherical particles with 212-355 $\mu$m diameter and the master alloy.
Figures \ref{fig:MCE}(a) and \ref{fig:MCE}(b) show
the magnetic entropy change $\Delta S_{\rm M}$
and the adiabatic temperature change $\Delta T_{\rm ad}$
for the magnetic field change $\mu_{0}\Delta H =$ 1, 3, and 5 T,
where the initial magnetic field was set to zero.
Both the $\Delta S_{\rm M}$ and the $\Delta T_{\rm ad}$ for each $\mu_{0}\Delta H$
take a maximum value at around $T_{\rm C}$.
For the spherical particles,
the maximum values for $\mu_{0}\Delta H =$ 5 T are
23.9 J kg$^{-1}$ K$^{-1}$ and 9.1 K, respectively,
which are close to those for the master alloy (25.7 J kg$^{-1}$ K$^{-1}$ and 9.7 K).
The values of $\Delta S_{\rm M}$ are consistent with those reported in the literatures
\cite{Campoy-RPB-2006,Khan-JAP-2011}.
Surprisingly, to our best knowledge, this is the first report on
$\Delta T_{\rm ad}$ for $\mu_{0}\Delta H =$ 5 T in HoAl$_{2}$.
Its maximum value is comparable to the reported value of 10 K for ErCo$_{2}$
\cite{Wada-JAC-2001},
a well-known giant magnetocaloric material with a similar $T_{\rm C}$ of 32 K.
For further comparisons, see the table in Ref. \cite{Pedro-NPG-2020}
summarizing magnetocaloric properties of various materials with $T_{\rm C}$ below 30 K.

It would be noticed that the shape of $\Delta S_{\rm M}$ and $\Delta T_{\rm ad}$ curves
below $T_{\rm C}$ is a bit different between the spherical particles and the master alloy.
Anisotropic MCE has been reported in HoAl$_{2}$ single crystal
in association with the spin-reorientation transition
\cite{Patra-JPCM-2014}.
In this context, Gil et al. \cite{Gil-JMMM-2016} have pointed out that
the texture effects in polycrystalline samples can affect
the shape of $\Delta S_{\rm M}$ especially for $T \leq$ 20 K.
Accordingly, sample dependence possibly appears,
but it will not be a serious problem for the performance of
the HoAl$_{2}$ gas-atomized particles as a magnetic refrigerant.
%
\section{Conclusions}
The electrode induction melting gas atomization process was successfully applied for
fabricating magnetocaloric particles of the Laves phase compound HoAl$_{2}$
by using the sintered rod as the electrode rod.
The particle size range of the resulting particles is suitable for AMR systems,
with the highest yield of 14-22 wt\% of the total atomized powder for 212-355 $\mu$m diameter.
These gas-atomized particles have mostly spherical shapes and few internal pores
and are also less contaminated during processing.
Furthermore, they exhibit the sharp ferromagnetic transition
and large magnetocaloric properties, similar to the bulk counterpart.

All these results indicate that the HoAl$_{2}$ gas-atomized particles are
of potential use as a magnetic refrigerant.
Nevertheless, we notice that other important properties should be guaranteed
for these particles before implemented in AMR systems.
In particular, mechanical stability is a significant factor
in determining the usefulness of the produced particles,
and we have a future plan to assess the mechanical properties of the gas-atomized particles.
Meanwhile, the technique devised in this study would facilitate
the stable production of spherical particles of various brittle magnetocaloric materials
with high melting points, including the RT$_{2}$ compounds,
helping develop high-performance magnetic liquefiers for hydrogen.
%
\section*{Acknowledgements}
This work was supported by JST-Mirai Program Grant Number JPMJMI18A3, Japan.
%

\newpage
%

\end{document}